\documentclass[se, manuscript]{copernicus}

\nolinenumbers

\begin{document}

\title{Parallel numerical simulation of impact crater with perfect matched layers}

\Author[a]{Huacheng}{Li}
\Author[b]{Zongyu}{Yue}
\Author[a][nan\_zhang@pku.edu.cn]{Nan}{Zhang}
\Author[c]{Jinhai}{Zhang}
\Author[c]{Zhongzheng}{Miao}

\affil[a]{Key Laboratory of Orogenic Belts and Crustal Evolution, School of Earth and Space
Sciences,Peking University, Beijing, 100871, China}

\affil[b]{Key Laboratory of Earth and Planetary Physics, Institute of Geology and
Geophysics,Chinese Academy of Sciences, Beijing, 100029, China}

\affil[c]{Engineering Laboratory for Deep Resources Equipment and Technology, Institute of
Geology and Geophysics,Chinese Academy of Sciences, Beijing, 100029, China}




\runningtitle{Parallel numerical simulation of impact crater with perfect matched layers}
\runningauthor{Li et al.}

\received{}
\pubdiscuss{} 
\revised{}
\accepted{}
\published{}


\firstpage{1}

\maketitle

\begin{abstract}
Impact craters are the primary geomorphic features on the surfaces of
celestial bodies such as the Moon, and their formation has significant im-
plications for the evolutionary history of the celestial body. The study of
the impact crater formation process relies mainly on numerical simulation
methods, with two-dimensional simulations capable of reproducing general patterns of impact processes while conserving computational resources. However, to mitigate the artificial reflections of shock waves at numerical boundaries, a common approach involves expanding the computational domain, greatly reducing the efficiency of numerical simulations. In this study, we developed a novel two-dimensional code SALEc-2D that employs the perfect matched layer (PML) method to suppress artificial reflections at numerical boundaries. This method enhances computational efficiency while ensuring reliable results. Additionally, we implemented MPI parallel algorithms in the new code to further improve computational efficiency. Simulations that would take over ten hours using the conventional iSALE-2D code can now be completed in less than half an hour using our code, SALEc-2D, on a standard computer. We anticipate that our code will find widespread application in numerical simulations of impact craters in the future.
\end{abstract}


\introduction  
Impact craters are one of the most common landforms on the solar solid bodies, and they are the products of high-speed meteorite impacts \citep{Melosh1989}. On the Moon, the high-speed impacting process can cause significant heating \citep{Kurosawa2018},  lead to the lunar mantle rebound \citep{Melosh2013},  excavate and  eject material from the depths to the surface \citep{Fassett2011}, etc. Thus the formation of impact craters has significant influence on the geologic evolution of the Moon. In studying the dynamics of impact cratering, computer simulations is a powerful tool and it is even the only feasible method for studying the formation of impact structures larger than about 1 km in diameter and for impact velocities typical of the solar system \citep{Pierazzo2008}. Currently there many hydrocodes available to simulate the impact cratering process, such as CTH \citep{Mcglaun1990}, iSALE-2D/3D \citep{Wunnemann2006,Elbeshausen2009}, SOVA \citep{Shuvalov1999}, SPH \citep{Benz1994,Benz1995}, SALEc \citep{Li2022}, etc.

In previous cratering simulations, the code iSALE-2D has been widely used \citep[e.g.,][]{Ogawa2021a,Prieur2018,Rajsic2021}.
Although the 3D hydrocode is necessary for the most probable impact angle of 45 degree to the horizontal \citep{Shoemaker1962}, the computational cost and required resources are still too expensive, which hence endows an irreplaceable value of 2D hydrocode.
2D simulations can provide substantially higher resolution and then a more accurate representation of vertical impacts than 3D simulations \citep{Potter2012}, and the oblique impact process can be estimated by the vertical velocity component\citep{Pierazzo2000,Elbeshausen2009,Elbeshausen2013}.
iSALE-2D is based on the SALE hydrocode solution algorithm \citep{Amsden1980}.
To simulate hypervelocity impact processes in solid materials,  an elasto-plastic constitutive model, fragmentation models, various equations of state (EoS), and multiple materials \citep{Melosh1992,Ivanov1997}
were included. In addition,  a modified strength model \citep{Collins2004}, a porosity compaction model \citep{Wunnemann2006,Collins2011a} and a dilatancy model \citep{Collins2014} were also incorporated. The iSALE-2D code has been have been benchmarked against other hydrocodes \citep{Pierazzo2008} and validated against experimental data from laboratory scale impacts \citep{Pierazzo2000,Davison2011,Miljkovic2013}. However, to suppress the influence of artificial reflections from the numerical boundaries,  a much larger model is usually necessary in iSALE-2D simulations. For example, in simulating the basin forming impacts, \cite{Rolf2017} used a model domain of at least 10 times of the basin diameter. This undoubtedly greatly reduces computational efficiency. 
 
To overcome those limitations of iSALE-2D, we developed a new hydrocode SALEc-2D for impact cratering simulations. In this code, a method of the perfect matched layer that is superior to many other methods \citep{Gao2017} is adopted to diminish the reflected shock wave from the numerical boundaries, thus a computational domain of a size comparable to the impact crater is enough for the simulations. In addition, we employ the MPI library in SALEc-2D, which further significantly increase the efficiency of the cratering simulations. It is expected that the SALEc-2D code would be widely applied in numerical simulations of impact craters in the future.

\section{Multiprocessing in SALEc-2D}

SALEc-2D is the two-dimensional version of SALEc \citep{Li2022}, 
and many modules in SALEc-2D (e.g. equation of state, VOF material surface, strength, and input/output) are inherited from SALEc except for the discretization of conservation equations and parallelism modules. Similar to SALEc, the algrithom of SALEc-2D is also based on that of SALE (Amsden et al., 1980), which split the conservation equations to Lagrangian and Eulerian steps. The Lagrangian step basically solves conservation equations of momentum and energy:
\begin{align}
        \rho \partial_t \vec v + \nabla \cdot (p - \tau) - \rho \vec g &= 0 \\
	\rho\partial_t I + p\nabla\cdot\vec v - \tau:\nabla\vec v &= 0
\end{align}
While the Eulerian step solves the advection equations of mass, momentum, and energy:
\begin{align}
    \partial_t\rho + \nabla\cdot(\rho \vec v) &= 0 \\
    \partial_t (\rho\vec v) + \nabla  \cdot (\rho \vec v  \vec v) &= 0 \\
    \partial_t  (\rho I) +  \nabla \cdot  (\rho \vec v I) &= 0 
\end{align}
where $\rho,\vec v, p,\tau, I, \vec g$ is the density, velocity, pressure, stress, and gravity respectively. An abstract computing flow of SALE is shown in Figure 1a.

To accelerate the simulation, SALEc-2D employs multiprocessing and decomposes the computation domain into equal-sized subdomains, with each subdomain processed by a single processor. All of the grids in a subdomain are divided to three types (Figure 1b):

(a) Receive section: They are the ghost grids that are at the margin of subdomains to hold data from its neighbour processors. In every cycle, qualities on ghost grids must be prepared before updating the corresponding quantities on its neighbour grids.

(b) Send section: They are the grids on which qualities will be sent to neighbour processors and stored in ghost grids of neighbour processors.

(c) Internal section: They are the grids on which updating qualities don't require any ghost grid data;

In the SALEc-2D code, the Message-Passing Interface (MPI) library is used to realize the above multiprocessing, and between the neighbour processors the non-blocking communication is used in each cycle. Figure 1c shows how quantities are updated in parallel environment. Updating parallel qualities start with posting the sending/receiving of ghost. At the same time, update qualities on internal section. After calculation of internal section, wait for completing of sending/receiving on ghost. Then update qualities on the send section.
\begin{figure}
    \centering
    \includegraphics[width=0.8\textwidth]{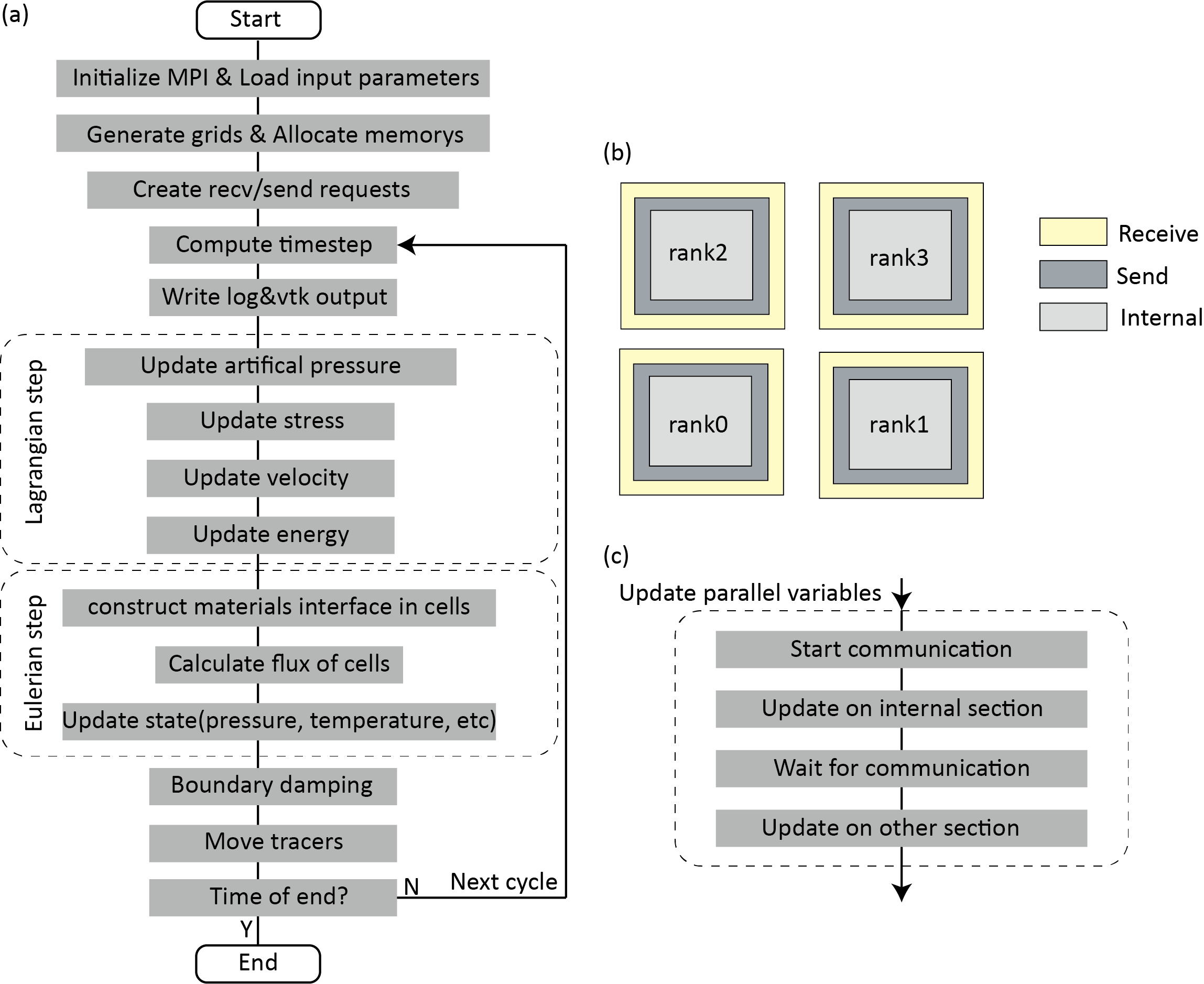}
    \caption{(a) Computing flow diagram of SALEc-2D, (b) Computation domain decomposing, (c) Updating of parallel variables with non-block communication}
    \label{fig:salec-flow}
\end{figure}

\section{PML method in SALEc-2D}
The PML method is commonly applied in wave equation\citep{Berenger1994}, which can effectively remove the reflected wave from the numerical boundary. The PML method basically extends the spatial partial derivative to complex coordinate \citep[i.e.][]{Johnson2021}:
\begin{equation}
    \frac{\partial }{\partial x} 
    \to 
    \frac{1}{1 + \frac{d(x)i}{\omega}}
    \frac{\partial }{\partial x} 
    \label{eq:pml-extend}
\end{equation}
where $d(x)$ is the attenuation coefficient of the PML, $\omega$ is the frequency and $i=\sqrt{-1}$.
In impact crater simulation, the Eulerian conservation equations are as follows: 
\begin{align}
\partial_t \rho +  \nabla \cdot (\rho\vec v) &= 0 \label{eq:ct0} \\
 \partial_t (\rho\vec v) + \nabla \cdot (\rho \vec v \vec v)  - \nabla\cdot(\tau-p + p_0) &= 0 \\
\partial_t (\rho I) + \nabla \cdot (\rho \vec v I)  + p\nabla\cdot\vec v- \tau : \nabla\vec v \label{eq:ct2} &= 0
\end{align}
where $p_0$ is the initial pressure before impact and is constrained by 
$-\nabla p_0 + \rho \vec g = 0$. 
In the region near numerical boundary, we assume the change of $\rho, \vec v$ and $I$ can be represented by a wave fluctuation term relative the initial state. For the time-harmonic solutions of eq.(\ref{eq:ct0}-\ref{eq:ct2}) with angular frequency $\omega$, substitute the spatial partial derivative of eq.\eqref{eq:pml-extend}
in y direction
\begin{align}
    -i\omega \rho + \partial_x (\rho v_x) + \frac{1}{1 + \frac{d_yi}{\omega}}\partial_y (\rho v_y)
    = 0\\
    -i\omega \rho v_x + \partial_x (\rho v_x v_x) + \frac{1}{1 + \frac{d_yi}{\omega}} \partial_y (\rho v_y v_x) 
    + \partial_x (p - \tau_{xx} -p_0) - \frac{1}{1 + \frac{d_yi}{\omega}}\partial_y\tau_{xy} &= 0 \\
	-i\omega \rho v_y +\partial_x (\rho v_x v_y) + \frac{1}{1 + \frac{d_yi}{\omega}} \partial_y (\rho v_y v_y) 
	- \partial_x\tau_{xy}+\frac{1}{1 + \frac{d_yi}{\omega}}\partial_y(p-\tau_{yy}-p_0) &= 0 \\
    -i\omega \rho I + p\partial_xv_x + p\frac{1}{1 + \frac{d_yi}{\omega}}\partial_yv_y 
    +\partial_x (\rho v_x I) + \frac{1}{1 + \frac{d_yi}{\omega}} \partial_y (\rho v_y I)
    & \nonumber\\
    - (\tau_{xx}\partial_xv_x + \tau_{xy}\partial_xv_y + \tau_{yy}\frac{1}{1 + \frac{d_yi}{\omega}}\partial_yv_y + \tau_{xy}\frac{1}{1 + \frac{d_yi}{\omega}}\partial_yv_x) &= 0 \label{eq:pml-eq2}
\end{align}
Introduce auxiliary variables $\phi_1,\phi_2,\phi_3,\psi_1,\psi_2$ such that:
\begin{align}
	\partial_t\phi_1 &= d_y \partial_x (\rho v_x) \\
	\partial_t\phi_2 &= d_y \partial_x (\rho \vec v v_x)  \\
	\partial_t\phi_3 &= d_y\partial_x(\rho v_x I)  \\
        \partial_t\psi_1 &= d_y\partial_x
	\begin{bmatrix}
		p - \tau_{xx} -p_0 \\
		-\tau_{xy}
	\end{bmatrix} \\
        \partial_t\psi_2 &= d_y(p-\tau_{xx})\partial_xv_x -d_y\tau_{xy}\partial_xv_y
\end{align}
Auxiliary variables for x direction have similar expressions:
\begin{align}
	\partial_t\phi_1 &= d_x \partial_y (\rho v_y) \\
	\partial_t\phi_2 &= d_x \partial_y (\rho \vec v v_y)  \\
	\partial_t\phi_3 &= d_x\partial_y(\rho v_y I)  \\
        \partial_t\psi_1 &= d_x\partial_x
	\begin{bmatrix}
        -\tau_{xy} \\
		p - \tau_{yy} -p_0 
	\end{bmatrix} \\
        \partial_t\psi_2 &= d_x(p-\tau_{yy})\partial_yv_y -d_x\tau_{xy}\partial_yv_x
\end{align}
The control equations in PML absorbing zone are:
\begin{align}
\partial_t \rho +  \nabla \cdot (\rho\vec v) &= -d_y\rho - \phi_1 \label{eq:pml-ct0} \\
 \partial_t (\rho\vec v) + \nabla \cdot (\rho \vec v \vec v)  - \nabla\cdot(\tau-p + p_0) &= -d_y\rho \vec v - \phi_2 -\psi_1\\
\partial_t (\rho I) + \nabla \cdot (\rho \vec v I)  + p\nabla\cdot\vec v- \tau : \nabla\vec v \label{eq:pml-ct2} &= -d_y\rho I - \phi_3 -\psi_2
\end{align}
The right-hand sides of eq.(\ref{eq:pml-ct0}-\ref{eq:pml-ct2}) represents terms absorbing  shock waves. In the PML absorbing zone, we insert the auxiliary variables $\psi_1,\psi_2$ into the Eulerian step and $\phi_1,\phi_2,\phi_3$ into Lagrangian step. In addition, the attenuation coefficients is obtained by \citep{Collino2001}:
\begin{align}
    d_y(y) = \frac{3c}{2\delta} (\frac{y}{\delta})^2 \log\frac{1}{R}
\end{align}
where $c$ is the maximum acoustic wave velocity, $\delta$ is the thickness of PML boundary,  $y$ is the distance from the exact boundary, and $R$ is the theoretical reflection coefficient. We use $R=0.01$ when $\delta=30dy$.

\section{Verification of the PML method in SALEc-2D}
We simulate a 14.4 km diameter dunite asteroid hitting a plane target vertically. The target is composed of three distinct layers, 2.8 km calcite representing the sediments, 30 km granite representing the crust, and a layer of  dunite over 150 km representing the mantle. This configuration approximates the condition for the formation of the Chicxulub crater\citep{Collins2008,Christeson2009}. The strength and acoustic fluid model parameters of different materials is detailed in Table \ref{tb:materials}. The numerical boundaries are set as freeslip for left/right, outflow for top, and noslip for bottom. Other parameters are documented in Table \ref{tb:model}.

\begin{figure}
    \centering
    \includegraphics[width=\textwidth]{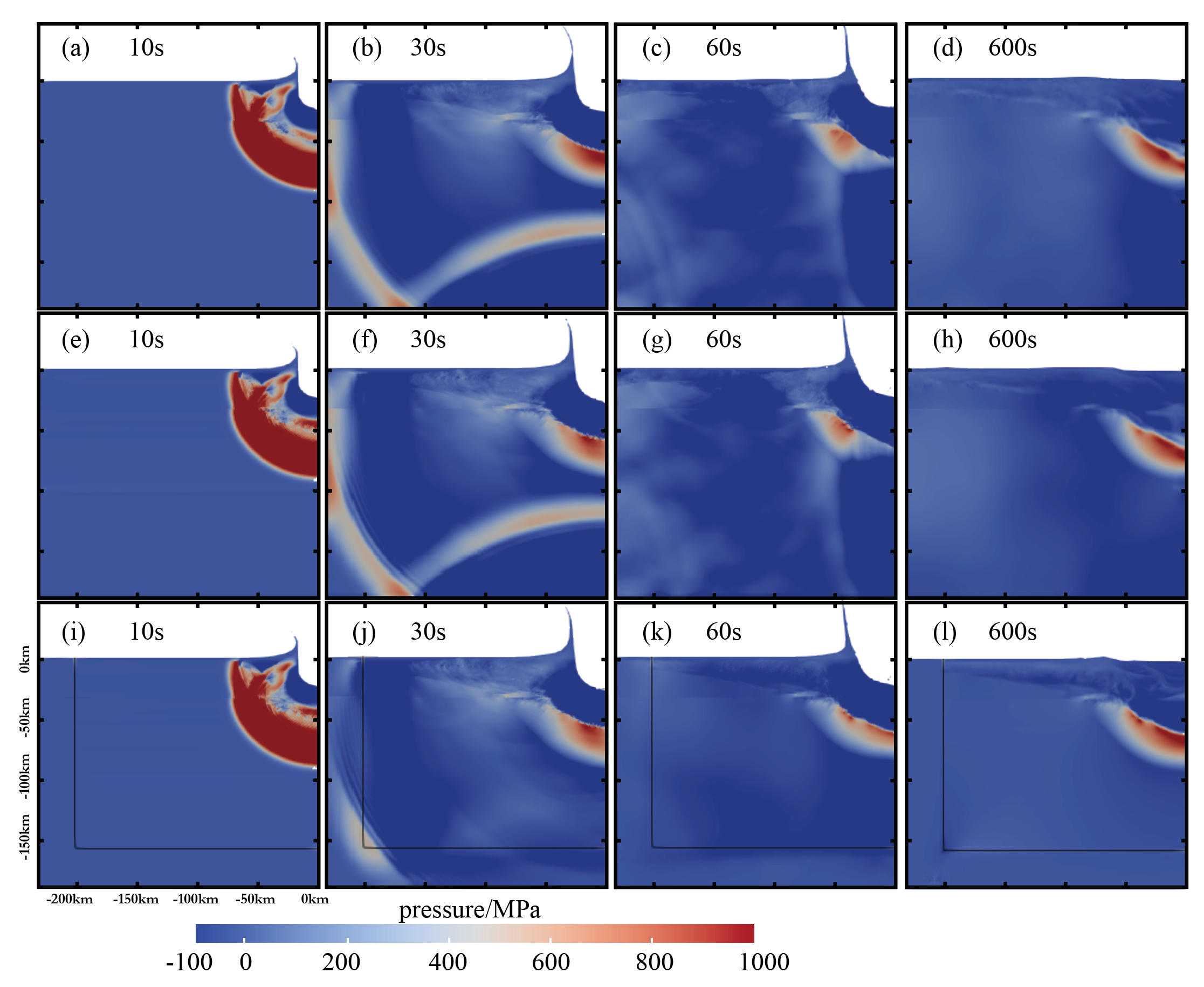}
    \caption{Pressure caused by impact predicted by iSALE-2D (a-d), SALEC-2D with out PML absorbing (e-h), and SALEC-2D with PML absorbing (i-l).
    The PML absorbing zone is put at the bottom and left.
    The black lines in (i-l) divide the calculation domain into the normal calculation area (X\textgreater -200 km and Y\textgreater -156 km) the PML absorbing boundary layers (X\textless-200 km or Y\textless-156 km).
    }
    \label{fig:pml-pressure}
\end{figure}

Figure \ref{fig:pml-pressure} illustrates the development of pressure (excluding the initial pressure) and the crater profile using different codes. The first row is from the iSALE-2D code, the central row is from the SALEc-2D code yet without using the PML method, and the bottom row is from the SALEc-2D code with the PML method. Each column represents the same time for the three calculations of 10 s, 30 s, 60 s, 600 s, respectively.
After 10 s of the projectile hitting, the maximum pressure reaches around 4.4 GPa, located 73 km from the impact point, and the three simulations provide almost the same results (Figure 2a, 2e, 2i). 
After 30 s, the shock wave pressure reaches 50 MPa and it has travelled about 230 km. 
In Figure 2b and 2f, the impact wave is approaching the left boundary and has been reflected back from the bottom. The reflected shock wave is obviously caused by the numerical boundary (The similarity between the results from iSALE-2D and our newly developed SALEc-2D without using PML method also proves the reliability of the SALEc-2D code). Clearly, the reflected shock wave has an artificial impact on the state of the target. In previous studies, usually a much wider model domain is used to diminish the reflected shock wave.
However, in Figure 2j, which uses the same model domain and only the PML method is applied in the boundary grids as shown outside the black lines in SALEc-2D, clearly shows that the shock wave attenuates gradually at the left boundary and  there is no obvious reflected wave from the bottom. 
This demonstrates that with the same model domain, the PML method can effectively suppressed the reflected shock wave from numerical boundaries, and then its effectiveness is greatly enhanced.
After 60 s, the reflected wave still have a magnitude of 5 MPa in Figure 2c and 2g.
Figure 2d, 2h, and 2l show the pressure in final craters. 
To analyze the influence caused by the reflected shock wave, the crater profile is detailed in Figure 3. In most of the time, the disparity of crater profile from
SALEc-2D, SALEc-2D+PML, and iSALE-2D is less than 1400 m (3.5 grid of calculate resolution), and the maximum difference appears around the crater rim.

\begin{table}[]
\caption{Material parameters}
\begin{tabular}{l|l|l|l}
\hline
\textbf{Constant} & \textbf{Granite} & \textbf{Dunite} & \textbf{Calcite} \\
\hline
$Y_{i0}$ Cohesion of intact material (MPa)& 10 & 10 & 5 \\ 
$\mu_i$ Coefficient of internal friction  &&&\\ 
for intact material & 2.0 & 1.2 & 1.0 \\ 
$Y_{im}$ Limiting strength at high pressure &&&\\
for intact material (GPa)& 2.5 & 3.5 & 0.5 \\ 
$Y_{d0}$ Cohesion of damaged material (MPa) & 0.01 & 0.01 & 0.01 \\ 
$\mu_{d}$ Coefficient of internal friction &&&\\
for damaged material & 0.60 & 0.60 & 0.40 \\ 
$Y_{dm}$ Limiting strength at high pressure &&&\\
for damaged material (GPa) & 2.5 & 3.5 & 0.5 \\ 
$\gamma_\eta$ dimensionless parameter &&&\\
for viscosity in block model & 0.008 & 0.08 & 0 \\ 
$\gamma_T$ dimensionless parameter &&&\\
for damping vibration in block model & 115 & 50 & 0 \\ 
$c_{vib}$ dimensionless ratio between &&&\\
vibration and material velocities & 0.1 & 0.1 & 0 \\ \hline
\end{tabular} \label{tb:materials}
\end{table}

\begin{table}[h]
\centering
\caption{Grid Parameters in Simulations}
\begin{tabular}{l|l|l|l}
\hline
\textbf{Constant} & \textbf{iSALE-2D} & \textbf{SALEc-2D} & \textbf{SALEc-2D+PML} \\
\hline
Resolution & 400 m& 400 m& 400 m\\
compution domain x(km)& $[0,231] $ & $[0,231]$ & $[0,231]$ \\
compution domian y(km)& $[-181,48.6]$ & $[-181,48.6]$ & $ [-181,48.6]$ \\
high-resolution zone $(n_x,n_y)^a$ & (462,462) & (462,462) & (462,462) \\
extension zone $(n_l,n_r,n_t,n_b)^b$ & (0,50,20,50) & (0,50,20,50) & (0,20,20,20) \\
PML zone $(p_r,p_b)^c$ & (0,0) & (0,0) & (30,30) \\
Max grid spacing & 1061 m& 1061 m& 1061 m\\
Grid extension factor & 1.05 & 1.05 & 1.05 \\
\hline
\end{tabular} \label{tb:model}
{

a: $n_x,n_y$ is the grids in x/y direction of high resolution zone.
b: $n_l,n_r,n_t,n_b$ is the number of extension grids at the left,right,bottom, and top side of the high resolution zone.
c: $p_r,p_b$ is the number of PML grids at the right and bottom side
}
\end{table}

\begin{figure}
    \centering
    \includegraphics[width=\textwidth]{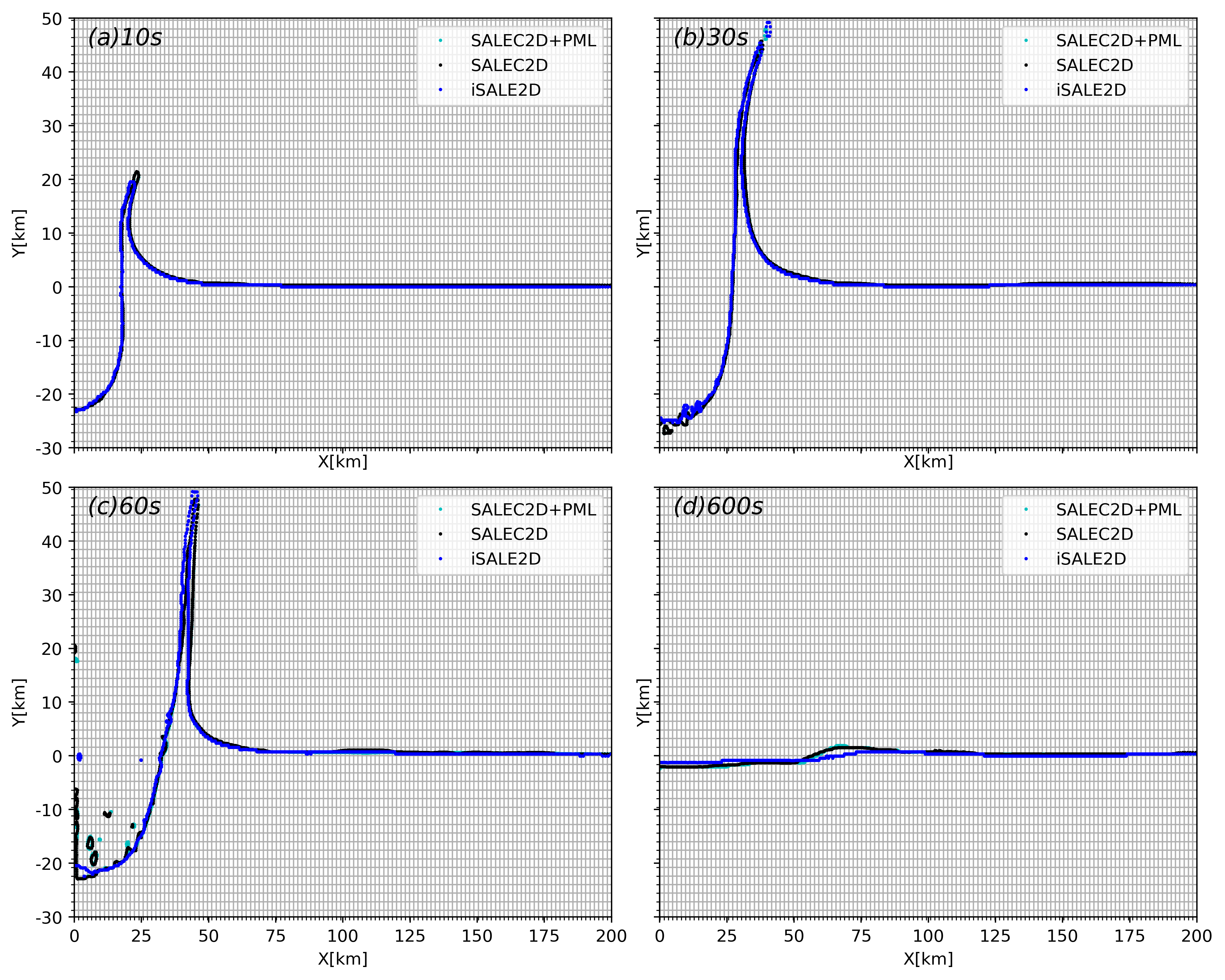}
    \caption{Crater profiles at 10 s, 30 s, 60 s, and 600 s. The gray grid-lines represent the grid with  1600 m resolution. SALEC2D+PML represents the results from SALEc-2D with the PML absorbing zone.
    SALEC2D is SALEc-2D without the PML absorbing zone.}
    \label{fig:crater-profile}
\end{figure}

\section{Influence of reflected wave: pressure anomaly and mantle deformation beneath final crater}
Although the reflected shock wave don't have too much effect to the final crater profile, it do have some artificial consequence on the simulations. The first is the pressure anomaly nearby the symmetry-axis. 
When the reflected wave approach symmetry-axis (X=0), it will reach a very high pressure for its energy  concentrates around the symmetry-axis.
In the Figure 2c or 2g, the reflected wave is only 5 Mpa, but when it is around the symmetry-axis at 56 s (Figure 4a or 4c), the pressure reaches 2.0 GPa.
After 104 s, 152 s, and 286 s of the impact, this pressure anomaly is still 1.0 GPa, 0.9 GPa and 0.7 GPa.
As a comparison, SALEc-2D+PML has eliminated the reflected wave effectively in the computation domain.
Except for the occasionally present high pressure, the area \textgreater 1 GPa predicted by iSALE2D is quite large than SALEc-2D+PML.
\begin{figure}[htp]
    \centering
    \includegraphics[width=\textwidth]{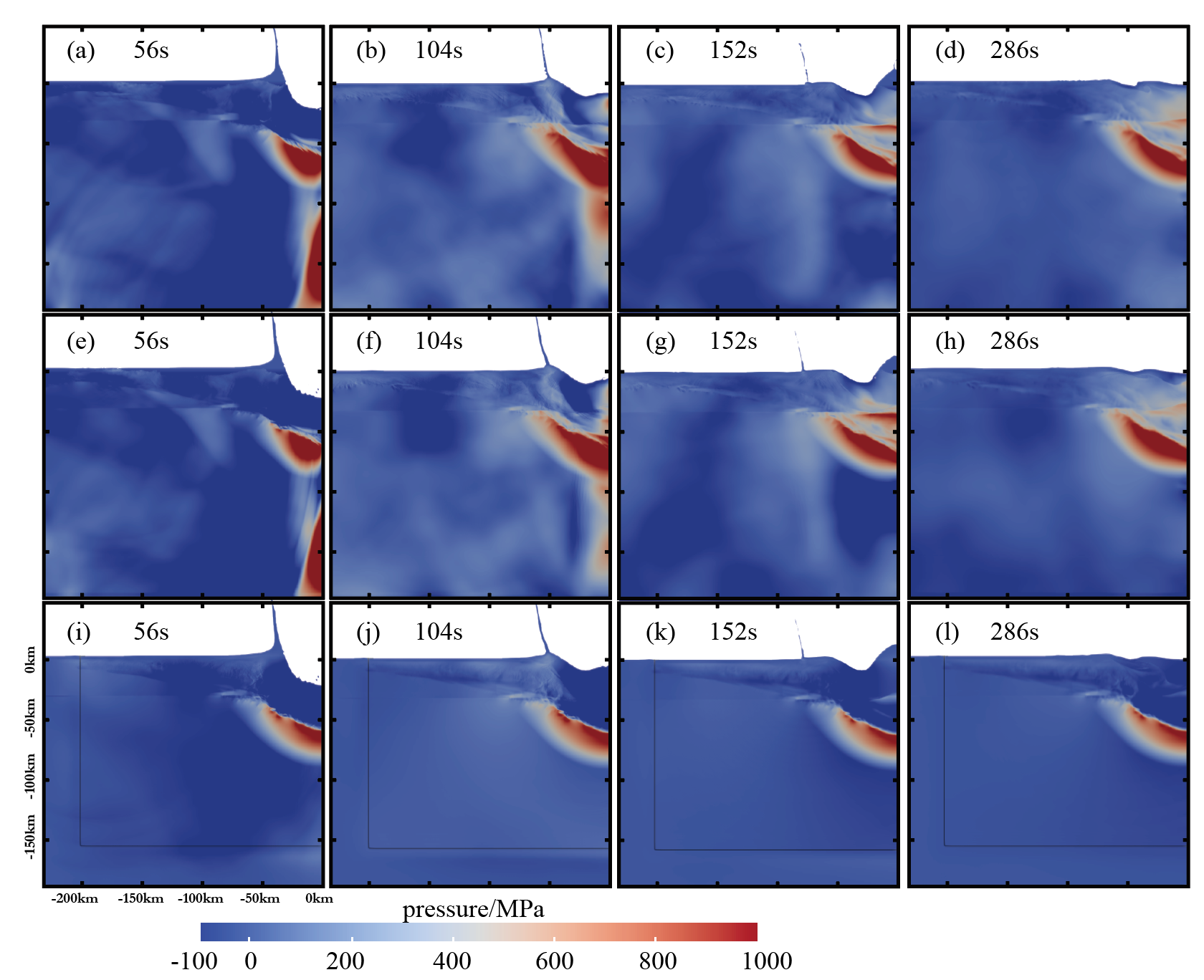}
    \caption{Pressure anomaly in simulation of iSALE-2D (a-d), SALEc-2D with out PML absorbing (e-h), and SALEc-2D with PML absorbing (i-l).}
    \label{fig:pressure-anomaly}
\end{figure}

Another consequence caused by the artificial reflected wave is the mantle beneath the crater. In Figure 5a, SALEc-2D and iSALE-2D give exact the same deformation of mantle.
The mantle is uplifted by 2.1 km near the center of final crater and depressed by 0.9 km at the distance of 32-42 km from the crater center. This is consistent with \cite{Christeson2009}.
When the PML boundary is applied, the uplift of mantle surface predicted by SALEc-2D+PML is 0.5 km higher than iSALE-2D. 
If the $\gamma_\eta$ of dunite is changed to 0.008, the uplift disparity between whether PML is utilized is \textgreater 4km (Figure 5b).
The influence of reflected wave on crust-mantle surface uplifting is related to the acoustic fluidization model.

\begin{figure}[htp]
    \centering
    \includegraphics[width=0.8\textwidth]{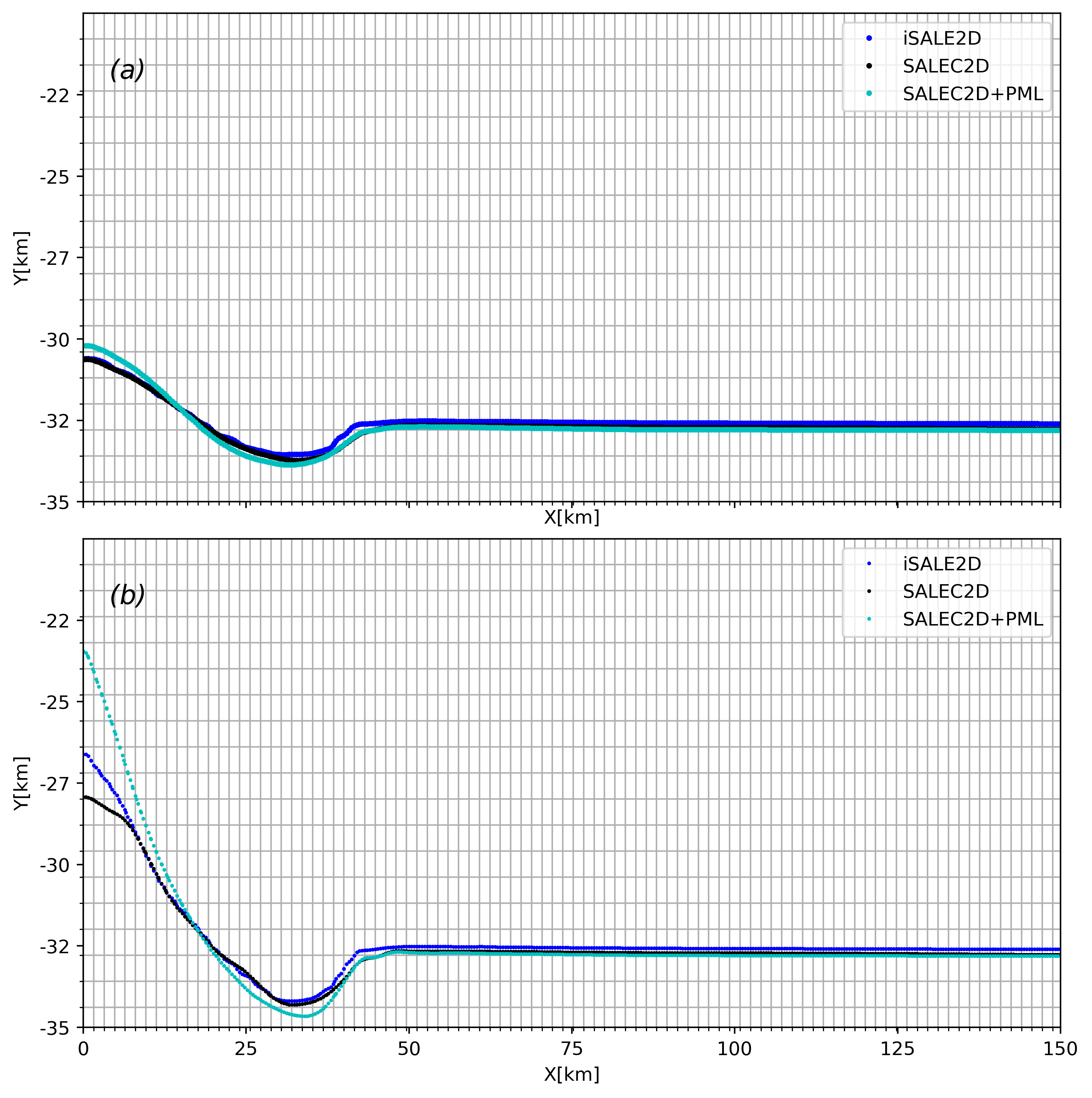}
    \caption{Uplift of crust-mantle surface. 
    iSALE2D,SALEC2D,SALEC2D+PML represent the mantle crust surface in the final crater
    predicted by iSALE-2D, SALEc-2D, and SALEc-2D with PML absorbing zone.
    (a) uplift in the final crater example with condition in Table 1. (b) the $\gamma_\eta$ of dunite is changed to 0.008.}
    \label{fig:uplift}
\end{figure}

\section{Parallel efficiency of SALEc-2D}

\begin{figure}
    \centering
    \includegraphics[width=0.8\textwidth]{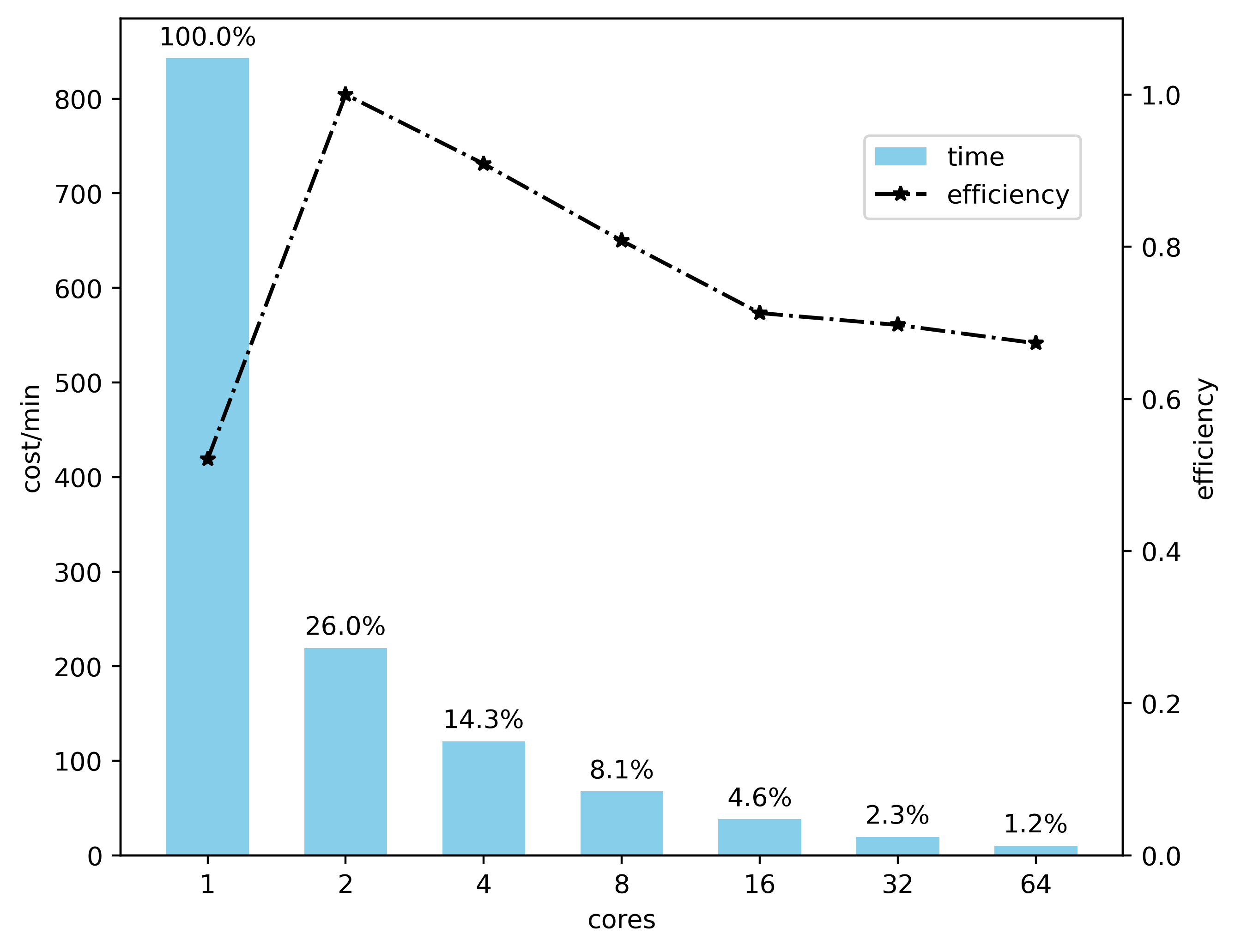}
    \caption{Time cost and efficiency with different cores running Chicxulub example. Single core result is from iSALE-2D. Cores 2-64 result is from SALEc-2D with PML absorbing.}
    \label{fig:salec-efficient}
\end{figure}

To test the parallel efficiency of SALEc-2D, we run the Chicxulub example with 1-64 cores on Intel(R) Xeon(R) CPU E5-2680 v4.
The time cost and parallel efficiency are show in Figure \ref{fig:salec-efficient}.
The single core code (iSALE-2D) used 842.1 min, while the parallel code (SALEc-2D) with 2 cores used 219.3 min (26\% of iSALE-2D). 
The time cost by SALEc-2D with 64 cores is only 10.2 min (1.2\% of iSALE-2D).
The parallel efficiency with n cores is defined as $P = \frac{2T_2}{nT_n}$ where $T_n$ is the time costed by n cores parallel code.
The efficiency of SALEc-2D drops to 0.67 on 64 cores because of the communication between processors, but still exceed that of iSALE-2D running on a single core.
It is noted that although both iSALE-2D and SALEc-2D utilize the SALE algorithm and finite volume method for discretizing the computation domain, even considering the external cost of PML zone in SALEc-2D, the efficiency of iSALE-2D is only half that of SALEc-2D.

\section{Stability of SALEc-2D}
\begin{figure}
    \centering
    \includegraphics[width=0.8\textwidth]{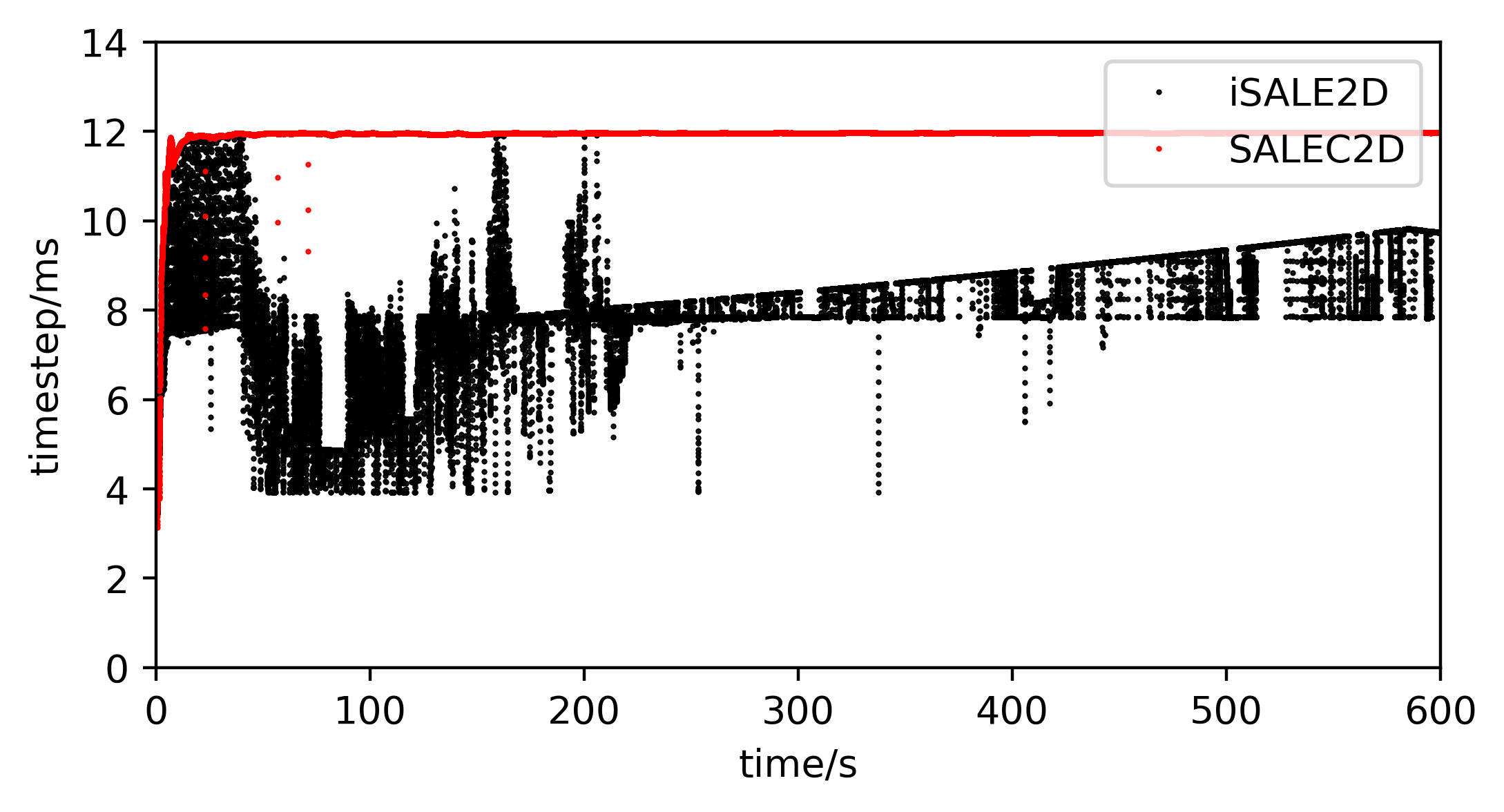}
    \caption{Timestep of SALEc-2D and iSALE-2D}
    \label{fig:timestep}
\end{figure}

\begin{figure}
    \centering
    \includegraphics[width=0.8\textwidth]{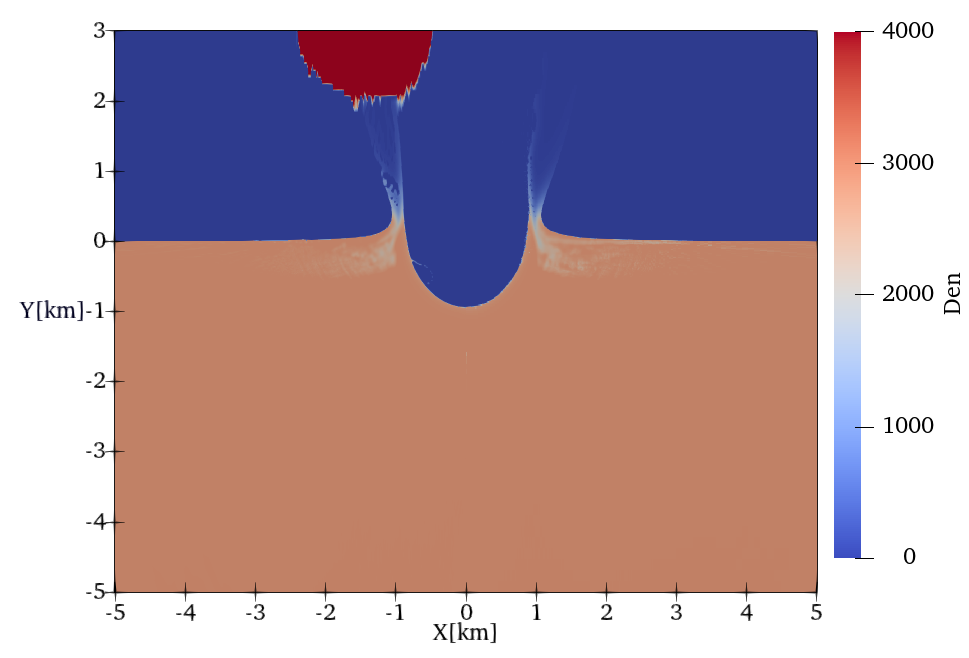}
    \caption{iSALE-2D crash example at time=2.3s, left $x<0$ represents the result of iSALE-2D, right $x>0$ represent the result of SALEc-2D. The detailed input for iSALE-2D is in the Supplement.}
    \label{fig:isale-crash}
\end{figure}

As an explicit hydrocode, to keep the solution stable, the timestep $dt$ of SALEc-2D is constrained by the Courant condition.
\begin{equation}
    dt \le C_0 \min\{ \frac{dx}{|\vec v|+c}, \frac{dy}{|\vec v|+c} \}
\end{equation}
where $dx,dy$ are the cell dimensions, $c$ is the speed of sound, and $C_0$ is a parameter $<1$ to limit the timestep. Both of the SALEc-2D and iSALE-2D set $C_0 = 0.2$ \citep{Amsden1980} and their timestep in the Chicxulub example is shown in Figure \ref{fig:timestep}.
The timestep of SALEc-2D only oscillates in the first 10s, whereas the oscillations of iSALE-2D persist throughout the entire simulation.
Those timestep oscillations in iSALE-2D may be attributed to some unidentified factors related to the code implementation. In iSALE-2D, following the Eulerian step, to prevent excessively small timestep, the surprise high velocity is deleted abruptly (the timestep of iSALE-2D is truncated around 4ms in Figure \ref{fig:timestep}). Beyond the timestep oscillations, iSALE-2D may unexpectedly crash under certain conditions (Figure \ref{fig:isale-crash}). This unstable scenario is also not observed in SALEc-2D.

\conclusions
In this study, we developed a new hydrocode, SALEc-2D, for numerical simulations of the impact crater formation process. The algorithm of this code is based on SALE, similar to the widely used iSALE-2D in current research on numerical simulations of impact craters. However, we added the PML method to this new code, which effectively suppresses the reflection of shock waves at numerical boundaries. As a result, the computational domain in numerical simulations only needs to be comparable in size to the impact crater being simulated. This significantly increases computational efficiency, as iSALE-2D typically requires the computational domain to be set 10 times the size of the impact crater to reduce the impact of reflected shock waves.

To validate the reliability of our code, we used the same input parameters as iSALE-2D and compared the simulation results. We found that when not using the PML absorbing boundary module in our developed SALEc-2D code, the results from both codes were nearly identical. Both exhibited strong reflected shock waves from the numerical boundaries, which have an impact on the pressure state experienced by the target and the structure of the underlying layers. In previous studies, these artificial reflected shock waves could only be eliminated by increasing the model space, which would greatly aggravate the computational overburden. In our SALEc-2D code, opening the PML absorbing boundary eliminates the reflected shock waves at the boundaries, producing reliable results even with a fairly small model. It is foreseeable that our SALEc-2D code will be widely applied in future research on numerical simulations of impact craters. The PML can be easily extended to 3D models, which would gain a much higher speed up ratio because of avoiding a great number of useless computations within the extended boundaries.


\codedataavailability{
The SALEc-2D code can be made available by the corresponding author on request. The iSALE-2D code release  is  distributed  on  a  case-by-case  basis  to  academic  users  in  the  impact  community,  strictly for non-commercial use. The input of iSALE-2D crashed sec
}


\videosupplement{
The input file and video of iSALE2D crashed example are available online at http://dx.doi.org/10.13140/RG.2.2.36277.45282
} 










\authorcontribution{
HL was responsible for the developing the SALEc-2D code and writing the original draft. NZ and ZY were responsible for the acquisition and management of financial support. JZ and ZM revised the PML algorithm in SALEc-2D. All authors contributed to editing of the manuscript.
} 

\competinginterests{The contact author has declared that none of the authors has any competing interests.} 


\begin{acknowledgements}
This work was supported by National Key Research and Development Program of China (2022YFF0503100), the B-type Strategic Priority Program of the Chinese Academy of Sciences (Grant No. XDB41000000),  National Natural Science Foundation of China (Grant No. 42241111), and the Key Research Program of the Institute of Geology and Geophysics, CAS (Grant No. IGGCAS-202204). We are grateful to the developers of iSALE-2D, including Gareth Collins, Kai Wünnemann, Dirk Elbeshausen, Tom Davison, Boris Ivanov and Jay Melosh.  Computations were conducted on the High-performance Computation Platform of Shannon Yun.
\end{acknowledgements}







\bibliographystyle{copernicus}
\bibliography{copernicus.bib}

\end{document}